# Elastically Buckled Film-Substrate System as a Two-dimensional Crystal

Wenqing Zhu[1,*]

[1] Department of Engineering, University of Cambridge, Cambridge CB2 1PZ, UK

[*]Email: wz379@cam.ac.uk

**Abstract**

Compressive mechanical stress exceeding a critical value leads to the formation of periodic surface buckling patterns in film-substrate systems. A comprehensive understanding of this buckling phenomenon is desired in applications where the surface topologies are modulated to achieve multifunctionalities. Here we reformulate the finite-deformation elastic theory of a film-substrate system by treating the compliant substrate as a nonlinear elastic solid. The resulting elastic free energy functional of the deflection field is shown to be equivalent to a minimal density functional of phase-field crystal theory plus a Gaussian curvature-related term. The proposed elastic model constructs a phase diagram based on free energy minimization, quantitatively agreeing with the buckling transitions observed in former experiments. The emerging hexagonal buckling system is shown to be equivalent to a two-dimensional crystal with proper scalings. We further conducted simulations of repeated buckling under cyclic stress to demonstrate a dynamically modulated structural adhesive, which resembles the physical process of repeated crystallization and melting near a critical temperature.

Crystalline systems with periodic structures, whether synthetic or formed in nature, have received intensive research interests given their simplicity and representativeness [1]. Understanding of their formation and evolution has significant implications in diverse fields such as condensed matter physics, engineering sciences, and life sciences [2-4]. Referring to the phase diagram of condensed matter, a disordered structure generally persists at a high temperature. It corresponds to a trivially uniform atomic density distribution on statistical average. When the matter is cooled down to a critical onset temperature, however, it crystallizes to form a periodic ordered structure minimizing the system free energy. Molecular dynamics simulations of crystallization with accurate descriptions of interatomic potentials have been able to provide details of the relevant thermodynamics and kinetics [5]. To probe the long-time behavior at a larger length scale, continuum models have also been proposed within the framework of dynamical density functional theory [6-9]. These models have been successfully applied to the studies of phase stability and defects dynamics (*e.g.*, the grain boundaries in graphene [10], which is a typical two-dimensional (2D) crystal with a hexagonal lattice).

A seemingly relevant procedure to crystallization is elastic buckling/wrinkling of soft matters [11]. Particularly, Cai *et al.* reported the emergence of a honeycomb hexagonal buckling pattern, similar to graphene lattice, on an elastic film-substrate system under equi-biaxial compressive stress [12]. This type of pattern formation is attributed to the system bifurcation to a minimized strain energy state with characteristic modes of out-of-plane deformation [13]. We argue that the similarity between crystallization and this elastic buckling lies in the fact that they share the common principle of free energy minimization, while a rigorous proof of the equivalence is yet to be given mathematically. A possible mapping from crystallization to elastic buckling implies that designated observations of elastic instability can be carried out on a film-substrate system to understand the thermodynamics of its mapped crystalline system, given proper parameters scaling. In addition, accurate predictions of surface buckling using a rigorous mathematical framework will enable the designs of dynamic pattern control, promoting wide applications in soft robotics with adjustable adhesion [14], sensing [15], measurements of elastic properties [16,17], *etc*.

In this Letter, we show that a buckled elastic film-substrate system is equivalent to a model 2D crystal according to the classical density functional theory developed by Elder *et al.* [6], except that an extra Gaussian curvature energy term needs to be considered for the elastic system. Examining the free energy of various periodic lattice phases analytically, we find that the Gaussian curvature-related energy contribution vanishes for a system larger than single unit cell, rendering the buckling phase diagram identical to that predicted by the density functional theory. Model validation and analyses of buckling characteristics are further presented, regarding the phenomenological crystallization temperature, lattice constants and crystalline defects.

We start by considering a Föppl-von Kármán film perfectly bonded on a nonlinear elastic substrate. A uniformly distributed equi-biaxial compressive stress $\sigma_0$ is applied on the film. The free energy of this film-substrate system reads $F = F_{\text{stretch}} + F_{\text{bend}} + F_{\text{sub}}$, where the stretching energy of the film

$$F_{\text{stretch}} = -\int d\bm{x} \left[ \nabla \cdot \bm{u} + |\nabla w|^2 \right] \sigma_0 h / 2, \tag{1}$$

and the bending energy of the film

$$F_{\text{bend}} = \frac{Eh^3}{12(1-\nu^2)} \int d\bm{x} \left[ (\nabla^2 w)^2 / 2 - (1-\nu) \det(\nabla\nabla w) \right]. \tag{2}$$

$\bm{u}$ and $w$ are the in-plane and out-of-plane displacement fields. $\nabla = \frac{\partial}{\partial x}\bm{e}_x + \frac{\partial}{\partial y}\bm{e}_y$ is the in-plane gradient operator. $E$ and $\nu$ are Young's modulus and Poisson's ratio of the film; $h$ is the film thickness. Assuming no shear deformation or relative sliding between film and substrate, the elastic energy stored in the substrate is written as

$$F_{sub} = \int d\boldsymbol{x} \left[ \frac{k_2}{2} w^2 + \frac{k_4}{4} w^4 \right]. \tag{3}$$

$k_2$ and $k_4$ are spring constants. The fourth-order term in $F_{sub}$ is introduced to describe the stiffening of polymeric substrates at large strains (see Supplemental Information). Therefore, the total free energy scaled by the bending modulus $B = Eh^3/\left[12(1-v^2)\right]$ reads

$$\tilde{F} = \frac{F}{B} = \int d\tilde{\boldsymbol{x}} \left[ \frac{\lambda}{4}\psi^4 + \frac{1-\varepsilon}{2}\psi^2 - |\tilde{\nabla}\psi|^2 + \frac{1}{2}(\tilde{\nabla}^2\psi)^2 - (1-v)\det(\tilde{\nabla}\tilde{\nabla}\psi) \right], \tag{4}$$

where a characteristic wavenumber $q = \sqrt{6\sigma_0(1-v^2)/(Eh^2)}$ is defined, such that $\tilde{\boldsymbol{x}} = q\boldsymbol{x}$ and $\psi = qw$. $\lambda = k_4/(Bq^6)$ and $\varepsilon = 1 - k_2/(Bq^4)$. Note that the $\boldsymbol{u}$ field-related term is not shown here, since we focus on the variation of energy functional with respect to $w$. Surprisingly, the expression of the scaled free energy in Eq.(4) is exactly the same as Swift-Hohenberg (SH) energy functional [18], except that there is an extra Gaussian curvature term $(1-v)\det(\tilde{\nabla}\tilde{\nabla}\psi)$. The phase-field crystal (PFC) model adopting SH functional has been successful in describing crystal nucleation and growth [6,9]. In this PFC model, the scaled free energy functional of atomic density reads

$$F_\rho = \int d\boldsymbol{x} \left[ \frac{\lambda_\rho}{4}\rho^4 + \frac{1-\varepsilon_\rho}{2}\rho^2 - |\nabla\rho|^2 + \frac{1}{2}(\nabla^2\rho)^2 \right], \tag{5}$$

where $\lambda_\rho$ and $\varepsilon_\rho$ are phenomenological parameters. Note that the increase in $\varepsilon_\rho$ reflects the decrease in temperature. The mathematical similarity between Eq. (4) and Eq. (5) suggests that the elastic film-substrate system studied herein can be treated as a model crystal, where the scaled deflection field $\psi$ is equivalent to a time-averaged atomic density field. Next, we check if the Gaussian curvature term in Eq. (4) influences the buckling phase selection. We calculate the scaled free energy densities of four different periodic phases, *i.e.*, the liquid phase (homogeneous phase) $\psi_l = \bar{\psi}$, the hexagonal lattice phase $\psi_h = \bar{\psi} + A_h \left[ \cos(q_h\tilde{x})\cos(q_h\tilde{y}/\sqrt{3}) + \cos(2q_h\tilde{y}/\sqrt{3})/2 \right]$, the square lattice phase $\psi_s = \bar{\psi} + A_s \cos(q_s\tilde{x})\cos(q_s\tilde{y})$, and the striped phase $\psi_t = \bar{\psi} + A_t \sin(q_t\tilde{x})$, respectively. In terms of the parameters $\lambda$, $\varepsilon$ and $\bar{\psi}$ (the spatial average of the scaled deflection), we have $f_l = \frac{1-\varepsilon}{2}\bar{\psi}^2 + \frac{\lambda}{4}\bar{\psi}^4$, $f_h = \frac{\lambda}{4}\bar{\psi}^4 + \frac{(1-\varepsilon)}{2}\bar{\psi}^2 + \frac{3(3\lambda\bar{\psi}^2 - \varepsilon)}{16}A_h^2 + \frac{3\lambda\bar{\psi}}{16}A_h^3 + \frac{45\lambda}{512}A_h^4$,

$f_s = \frac{\lambda}{4}\bar{\psi}^4 + \frac{(1-\varepsilon)}{2}\bar{\psi}^2 + \frac{3\lambda\bar{\psi}^2 - \varepsilon}{8}A_s^2 + \frac{9\lambda}{256}A_s^4$ and $f_t = \frac{\lambda}{4}\bar{\psi}^4 + \frac{(1-\varepsilon)}{2}\bar{\psi}^2 + \frac{3\lambda\bar{\psi}^2 - \varepsilon}{4}A_t^2 + \frac{3\lambda}{32}A_t^4$,

respectively. $A_h$, $A_s$ and $A_t$, are functions of $\lambda$, $\varepsilon$ and $\bar{\psi}$ (see Supplemental Information for detailed mathematical derivations). The unique third-order term of $A_h$ in $f_h$ is crucial for the stability of the hexagonal phase based on energy minimization. Importantly, we find that the Gaussian curvature term does not contribute to the total free energy, *i.e.*, $\int \det(\tilde{\nabla}\tilde{\nabla}\psi) d\tilde{\boldsymbol{x}} = 0$. It suggests that the expressions of free energy densities for all the four phases are identical to those derived from the PFC model in 2D [6]. Thus, obtained from the phase selection of energy minimization, the phase diagram of the elastic film-substrate system is the same as that derived from the 2D PFC model with SH functional (Eq. (5)) [6]. At high phenomenological temperature ($-\varepsilon$ is high and $\sigma_0$ is small), the system is in the liquid phase (buckle-free state). Decreasing the temperature (increasing $\sigma_0$) results in the emergence of hexagonal crystal phase. Further decreasing the temperature leads to the stripe phase, as shown in FIG. 1(a), where $\lambda = 1.1 \times 10^{-6}$ is used based on the materials parameters of the PDMS film-substrate system (see Supplemental Information) and $\sigma_0$ = 1 MPa. The theoretical $\varepsilon$-$\bar{\psi}$ phase diagram can be converted into a single-variable ($\sigma_0$) controlled curve (FIG. 1(b)). As the compressive stress $\sigma_0$ increases to the critical value $\sigma_{0,c}$, a hexagonal buckling pattern emerges. Further increasing $\sigma_0$ to $\sigma_{0,c}^{\text{Striped}} \approx 1.17 \sigma_{0,c}$ gives the striped buckling patterns formation. Previous theoretical analysis of elastic buckling predicted the formation of checkerboard square lattice phase in a film-substrate system under equi-biaxial stress, which contradicts the experimental observations [19]. In comparison, the present theory with a generalized SH functional successfully predicts the hexagonal phase over the metastable square lattice phase, agreeing with the experiment [12,19]. The critical stress for the onset of hexagonal phase is shown to increase with increasing the film modulus (with fixed $h$ = 0.01$H$) and the thickness (with fixed $E$ = 10 GPa) (FIG. 1(c) and (d)). Linear perturbation analysis by Chen and Hutchinson [20] gave the same trend, although the derived scaling law is different. Nevertheless, the present theory suggests that the ratio of critical stress of striped phase to that of hexagonal phase, $\sigma_{0,c}^{\text{Striped}}/\sigma_{0,c}$, remains a constant for a variety of thin film materials with different thickness or elastic moduli. This criterion can guide the controllable formation of hexagonal wrinkles in different film-substrate systems. Discussions on the potential applications of controllable pattern formation will be given later in the main text.

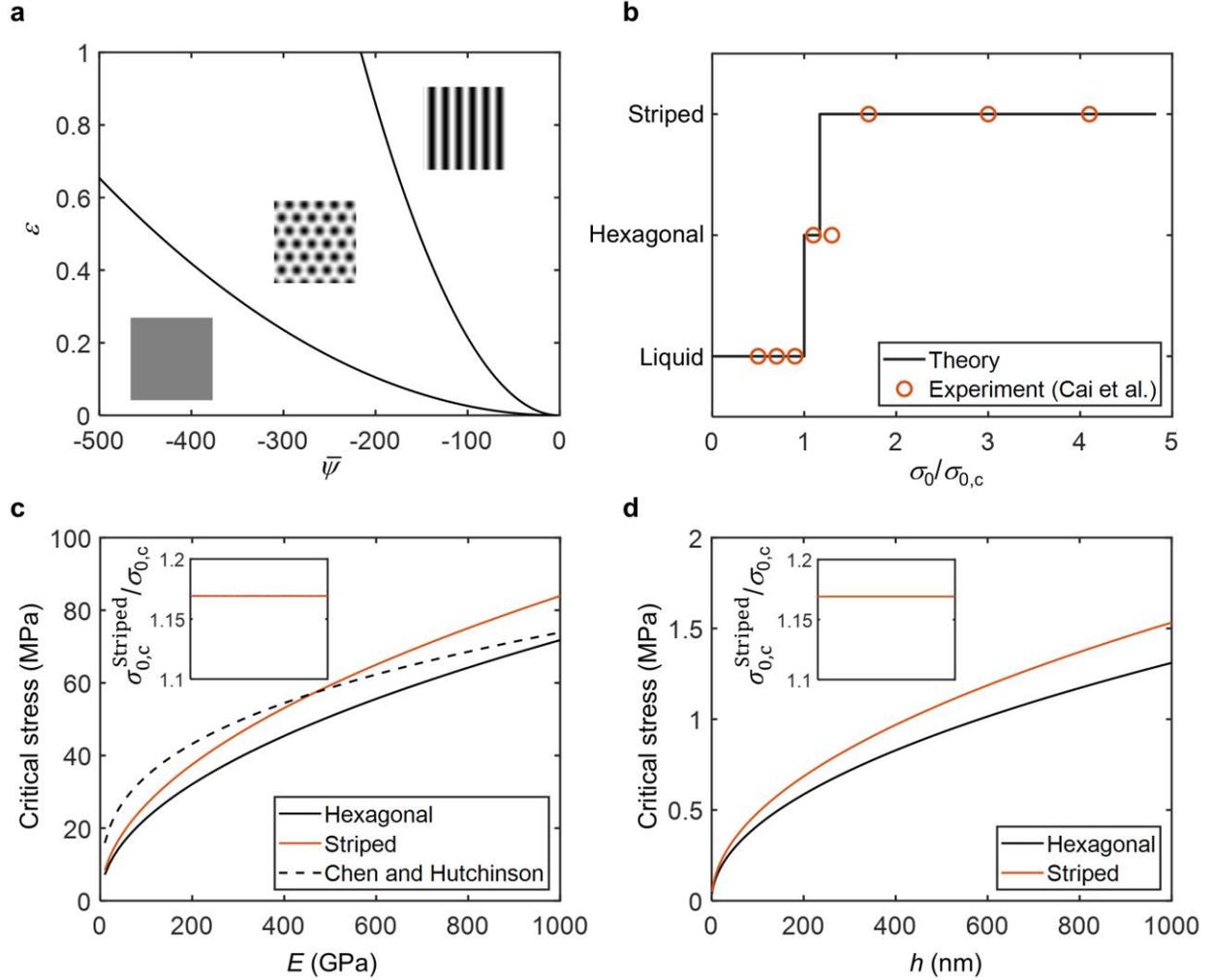

FIG. 1. Phase diagram of the elastic film-substrate system identical to that from the PFC model. (a) Phase selection for varying $\bar{\psi}$ and $\varepsilon$ at $\lambda = 1.1\times10^{-6}$. (b) Phase selection for varying the overstress $\sigma_0/\sigma_{0,c}$ from previous experiment (Ref. [12]) and theoretical prediction. Geometric and material parameters are adopted from Ref. [12]. (c-d) Critical compressive stress for the formation of hexagonal and striped phase versus $E$ and $h$. Dashed line is the prediction from the elastic buckling theory (Ref. [20]). Insets show the ratio of critical stresses.

Besides the phase diagram and the critical stress, it is necessary to delve into the structural characteristics of the buckling patterns. We focus on the regime of hexagonal lattice analogous to a 2D crystalline system. Numerical simulations are carried out based on the free energy functional (Eq. (4)) and a relaxational dynamics model (see Supplemental Information). Temporal evolutions of the $\psi$ field are monitored at different overstresses (different phenomenological temperatures). At a stress slightly higher than the critical value, i.e., $\sigma_0/\sigma_{0,c} = 1.01$ (temperature slightly below the solidification temperature), defect-free hexagonal lattice forms with uniform orientation (FIG. 2(a)). The corresponding fast Fourier transform (FFT) mapping of the entire domain also supports the hexagonal ordering (FIG. 2(c)). In comparison, increasing the overstress ($\sigma_0/\sigma_{0,c} = 1.05$) leads to a polycrystalline morphology with grain boundary defects (FIG. 2(b)). This aligns with the experimental observation of polycrystal-like buckling pattern at a deep buckling stage ($\sigma_0/\sigma_{0,c} \approx 1.3$), as shown in FIG. 2(d). The stress-dependent buckling pattern formation manifests an equivalent temperature-dependent solidification process. When a system is cooled down deeply to a low temperature, a large number of precursor crystal nuclei emerge and grow, as the crystallization condition is satisfied throughout the system. Since the nuclei orientations generally exhibit a random distribution, it facilitates a polycrystalline structure with random orientation that forms after grain coarsening. Moreover, we evaluate the lattice constant, which is defined as $\sqrt{3}$ times the side length of the hexagon. Without any adjustable parameter, the simulated lattice constant is in excellent agreement with the experiment, as can be seen from the matching peak position of the FFT intensity (FIG. 2(e)). In fact, the proposed theory also gives a consistent prediction: Energy minimization of a hexagonal lattice based on Eq. (4) gives $q_h = \sqrt{3}/2$ (see Supplemental Information). Thus, the lattice constant equals $2\pi/(q_h q) \approx 51$ μm, given that the characteristic wavenumber $q \approx 0.14$ μm$^{-1}$ for $\sigma_{0,c} < \sigma_0 < \sigma_{0,c}^{\text{Striped}}$.

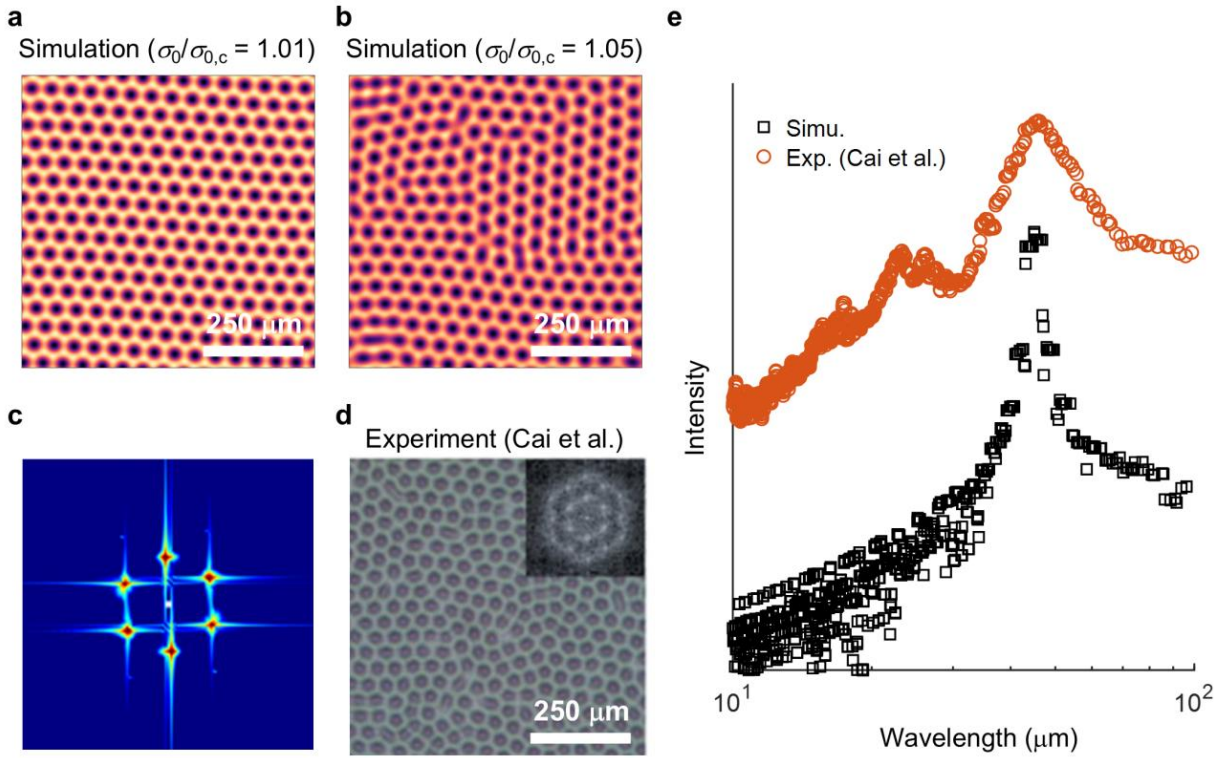

FIG. 2. Simulations of hexagonal pattern formation under constant compressive stress. (a-b) $\psi$ field at the overstress $\sigma_0/\sigma_{0,c} = 1.01$ and 1.05, respectively. (c) The corresponding FFT of the $\psi$ field at $\sigma_0/\sigma_{0,c} = 1.01$. (d) Experimental buckling pattern adopted from previous study (Ref. [12]). Inset shows the corresponding FFT. (e) Intensity of the FFT versus the wavelength. The characteristic wavelength (lattice constant) ≈ 45 μm of the hexagonal buckling pattern is identified as the peak position.

The buckling behavior resembling the crystallization of hexagonal lattice is crucial for modulating the surface properties of the film-substrate system. Here we propose a design of metasurface with tunable adhesive ability via controlling the magnitude of equi-biaxial compressive stress in the elastic film. This can be achieved by introducing cooling-induced thermal strain (FIG. 3(a)). Given the temperature drop $\Delta T$, the compressive stress $\sigma_0 = (\alpha_{sub} - \alpha_f)\Delta T/E$; $\alpha_f$ and $\alpha_{sub}$ are the thermal expansion coefficients of film and substrate, respectively. Inputting $\alpha_f = 10^{-5}$–$10^{-4}$ K$^{-1}$ for metal films and $\alpha_{sub} = 2.5\times10^{-4}$ K$^{-1}$ for PDMS [21], the compressive stress is around 20 MPa using a small temperature drop of 1 K, suitable to induce surface buckling. In a pick-and-place task of object movement, a hexagonal surface buckling pattern can significantly enhance the surface adhesive force $F_a$, as inspired by tree frog's adhesive pads [22], enabling the lifting of an object. When back to room temperature, the buckling pattern vanishes and $F_a$ decreases, releasing the object. Smart adhesion has also been achieved by heating to modify the polymeric material properties [23,24], but it requires a larger temperature change to be above the glass transition temperature, which is not energy cost-effective. The design proposed herein requires subtle temperature variation and does not introduce large hysteresis among pick-and-place cycling. We perform simulations of repeated buckling with cyclic compressive stress. During each cycle, when the stress magnitude exceeds the critical value, a polycrystalline pattern emerges (FIG. 3(b)). When the stress magnitude is below the critical stress, an unbuckled liquid state is retrieved (FIG. 3(c)). Note that the emerging polycrystalline structure at each cycle varies, having entirely different lattice orientation and grain boundary defects. It suggests the absence of memory effect for the unbuckled-buckled cycling, which is also the case for melting and recrystallization of a solid. Finally, it is worth stating that the physical temperature drop in the film-substrate system indeed leads to crystallization, in analogy to crystallization by cooling from high temperature for a condensed matter, despite that they hold different temperature scaling and critical onset temperatures.

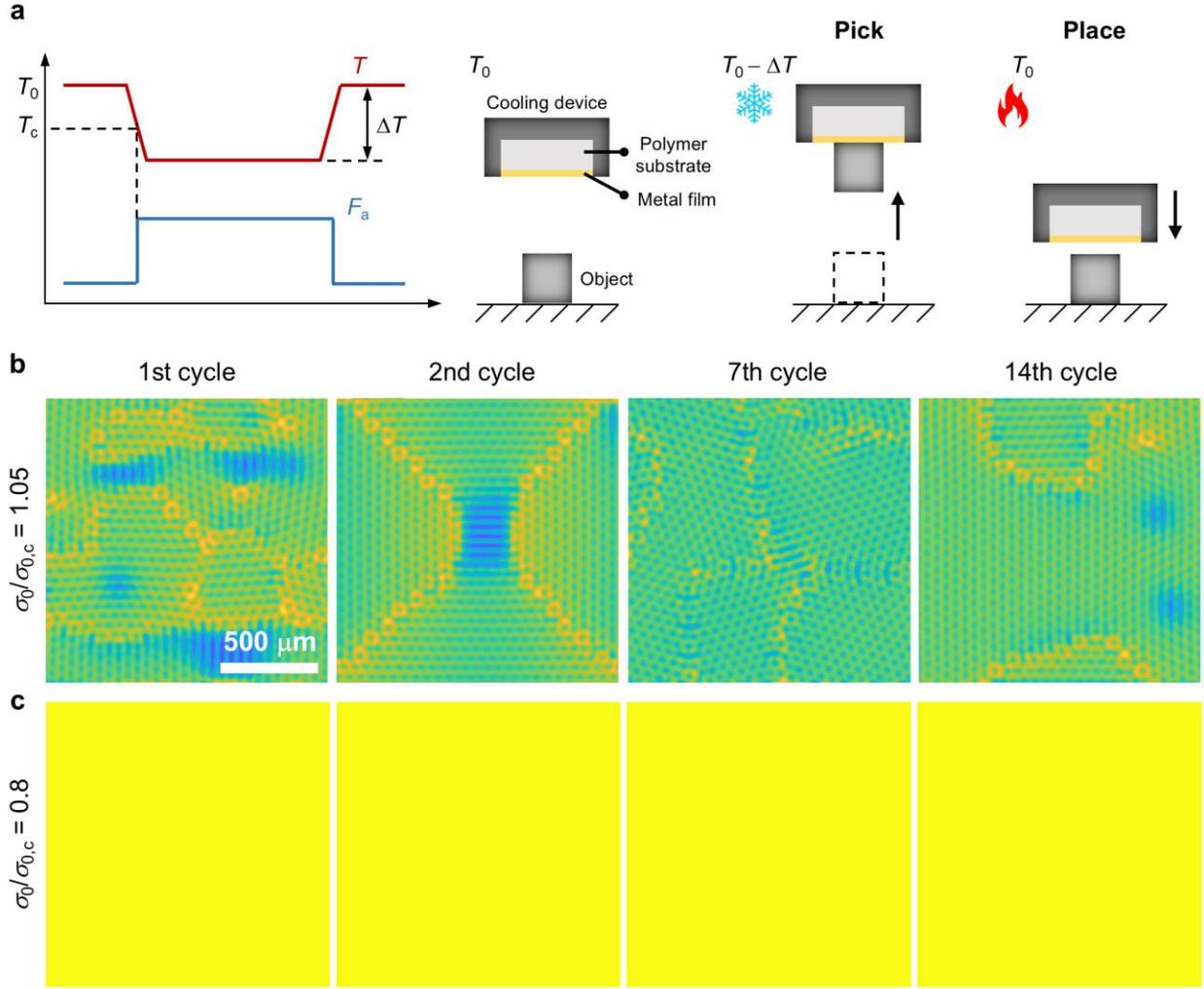

FIG. 3. Dynamic buckling (crystallization) for smart adhesive applications. (a) Schematic of pick-and-place manipulation with the adhesive force modulated by temperature. (b, c) Smoothed free energy density maps at $\sigma_0/\sigma_{0,c} = 1.05$ and $0.8$ for different loading cycles showing grain boundaries of the 2D crystal.

In summary, we prove that an equi-biaxially compressed elastic film-substrate system is equivalent to a model 2D crystal and exhibits a standard phase diagram predicted by the dynamical density functional theory. The proposed theoretical model clarifies the selection of hexagonal buckled state at a small overstress, which has not been able to be predicted by linear perturbation analysis. Importantly, the elastic film-substrate system can serve as a model crystal for understanding the history-dependent structural evolution. Using time-resolved and non-destructive optical approaches [25], it is expected to uncover detailed crystallization processes and defects dynamics. This is somehow more challenging to be revealed by in-situ observation of crystalline materials under electron microscopy, given the ultrashort timescale of atomic motion and influence of electron irradiation. These findings provides deeper understandings of elastic instability and offer a pathway to establishing an unified physical interpretation between crystallization and diverse phenomena of pattern formation such as Turing morphogenesis and hydrodynamic convection [13].

## Acknowledgements

This work has been supported by the National Natural Science Foundation of China (Grant No. 12302077).

# Supplemental Material for
# Elastically Buckled Film-substrate System as a Two-dimensional Crystal

Wenqing Zhu[1,*]

[1] Department of Engineering, University of Cambridge, Cambridge CB2 1PZ, UK

*Email: wz379@cam.ac.uk

## I. Free energy densities of different phases

Here we derive the mathematical expressions of scaled free energy densities. For the homogeneous liquid phase, there is no spatial variation of $\psi$. Thus,

$$f_l = \frac{1-\varepsilon}{2}\bar{\psi}^2 + \frac{\lambda}{4}\bar{\psi}^4. \tag{1}$$

For the hexagonal phase $\psi_h = \bar{\psi}_h + A_h\left[\cos(q_h\tilde{x})\cos(q_h\tilde{y}/\sqrt{3}) + \cos(2q_h\tilde{y}/\sqrt{3})/2\right]$, the scaled free energy of one unit cell (the unit cell area $S_h = 4\sqrt{3}\pi^2/q_h^2$) is

$$\begin{aligned}
f_h S_h &= \int_0^{\frac{2\pi}{q_h}} d\tilde{x} \int_0^{\frac{6\pi}{\sqrt{3}q_h}} d\tilde{y} \left[\frac{\lambda}{4}\psi_h^4 + \frac{1-\varepsilon}{2}\psi_h^2 - |\tilde{\nabla}\psi_h|^2 + \frac{1}{2}(\tilde{\nabla}^2\psi_h)^2 - (1-\nu)\det(\tilde{\nabla}\tilde{\nabla}\psi_h)\right] \\
&= \frac{\sqrt{3}\pi^2\lambda}{q_h^2}\bar{\psi}^4 + \frac{2\sqrt{3}\pi^2(1-\varepsilon)}{q_h^2}\bar{\psi}^2 + \left[\frac{3\sqrt{3}\pi^2(3\lambda\bar{\psi}^2+1-\varepsilon)}{4q_h^2} - 2\sqrt{3}\pi^2 + \frac{4\pi^2 q_h^2}{\sqrt{3}}\right]A_h^2 \\
&\quad + \frac{3\sqrt{3}\pi^2\lambda\bar{\psi}}{4q_h^2}A_h^3 + \frac{45\sqrt{3}\pi^2\lambda}{128q_h^2}A_h^4
\end{aligned} \tag{2}$$

Minimizing the free energy with $\partial f_h/\partial q_h = 0$ gives $q_h = \sqrt{3}/2$. Besides, considering $\bar{\psi} < 0$ without loss of generality, $\partial f_h/\partial A_h = 0$ gives $A_h = -4\bar{\psi}/5 + 4\sqrt{15\lambda\varepsilon - 36\lambda^2\bar{\psi}^2}/(15\lambda)$ [1]. Thus, we have

$$f_h = \frac{\lambda}{4}\bar{\psi}^4 + \frac{(1-\varepsilon)}{2}\bar{\psi}^2 + \frac{3(3\lambda\bar{\psi}^2 - \varepsilon)}{16}A_h^2 + \frac{3\lambda\bar{\psi}}{16}A_h^3 + \frac{45\lambda}{512}A_h^4. \tag{3}$$

For the square phase $\psi_s = \bar{\psi} + A_s\cos(q_s\tilde{x})\cos(q_s\tilde{y})$, the scaled free energy density ($S_s = 4\pi^2/q_s^2$) is

$$f_s = \frac{1}{S_s} \int_0^{\frac{2\pi}{q_s}} d\tilde{x} \int_0^{\frac{2\pi}{q_s}} d\tilde{y} \left[ \frac{\lambda}{4} \psi_s^4 + \frac{1-\varepsilon}{2} \psi_s^2 - \left|\tilde{\nabla} \psi_s\right|^2 + \frac{1}{2}\left(\tilde{\nabla}^2 \psi_s\right)^2 - (1-\nu)\det\left(\tilde{\nabla}\tilde{\nabla} \psi_s\right) \right]$$

$$= \frac{\lambda}{256}\left(9A_s^4 + 96\bar{\psi}^2 A_s^2 + 64\bar{\psi}^4\right) + \frac{1-\varepsilon}{8}\left(A_s^2 + 4\bar{\psi}^2\right) - \frac{q_s^2}{2} A_s^2 + \frac{q_c^4}{2} A_s^4 \quad (4)$$

Minimizing the free energy with $\partial f_s/\partial q_s = 0$ gives $q_s = \sqrt{2}/2$. Besides, $\partial f_s/\partial A_s = 0$ gives $A_s = 4\sqrt{(\varepsilon - 3\lambda\bar{\psi}^2)/\lambda}/3$. Thus, we have

$$f_s = \frac{\lambda}{4}\bar{\psi}^4 + \frac{(1-\varepsilon)}{2}\bar{\psi}^2 + \frac{3\lambda\bar{\psi}^2 - \varepsilon}{8} A_s^2 + \frac{9\lambda}{256} A_s^4. \quad (5)$$

For the striped phase $\psi_t = \bar{\psi} + A_t \sin(q_t \tilde{x})$, the scaled free energy density is

$$f_t = \frac{q_t}{2\pi} \int_0^{\frac{2\pi}{q_t}} d\tilde{x} \left[ \frac{\lambda}{4}\psi_t^4 + \frac{1-\varepsilon}{2}\psi_t^2 - \left(\partial_{\tilde{x}}\psi_t\right)^2 + \frac{1}{2}\left(\partial_{\tilde{x}\tilde{x}}^2\psi_t\right)^2 \right]$$

$$= \frac{\lambda}{32}\left(3A_t^4 + 24\bar{\psi}^2 A_t^2 + 8\bar{\psi}^4\right) + \frac{1-\varepsilon}{4}\left(A_t^2 + 2\bar{\psi}^2\right) - \frac{q_t^2}{2} A_t^2 + \frac{q_t^4}{4} A_t^4 \quad (6)$$

Minimizing the free energy with $\partial f_t/\partial q_t = 0$ gives $q_t = 1$. $\partial f_t/\partial A_t = 0$ gives $A_t = 2\sqrt{(\varepsilon - 3\lambda\bar{\psi}^2)/(3\lambda)}$. Thus, we have

$$f_t = \frac{\lambda}{4}\bar{\psi}^4 + \frac{(1-\varepsilon)}{2}\bar{\psi}^2 + \frac{3\lambda\bar{\psi}^2 - \varepsilon}{4} A_t^2 + \frac{3\lambda}{32} A_t^4. \quad (7)$$

For all these phases, it is found that the Gaussian curvature term does not contribute to the free energies, i.e., the integral over one unit cell $\int \det(\tilde{\nabla}\tilde{\nabla}\psi) d\tilde{x} = 0$. Therefore, from a thermodynamic perspective of energy minimization, the elastic film-substrate system is equivalent to the 2D phase-field crystal proposed by Elder et al. [2].

## II. Numerical simulations

Temporal evolution of the deflection field $\psi$ is simulated using finite element method. A conserved relaxational dynamics is adopted:

$$\frac{\partial \psi}{\partial t} = \tilde{\nabla}^2 \left(\frac{\delta \tilde{F}}{\delta \psi}\right) + \eta(\tilde{x}, t), \quad (8)$$

where $\tilde{F}$ is the scaled free energy functional as in Eq. (4) in the main text. The white noise term $\eta(\tilde{x},t) \sim N(0, \sigma_\eta^2)$ with $\sigma_\eta$ = 0.1-10. The initial values of $\psi$ are also sampled from a normal distribution, i.e., $\psi(\tilde{x}, 0) \sim N(\bar{\psi}, 1)$. The mesh size $\Delta \tilde{x} = \Delta \tilde{y} = \pi/4$. Periodic boundary conditions are employed. The input parameters are obtained from the previous work by Cai et al. [3]: $H$ = 3 mm, $h$ = 100 nm, $E$ = 10 GPa, $\nu$ = 0.5, $k_2$ = 0.37 Nm$^{-3}$ and $k_4$ = 0.11 Nm$^{-5}$. The spring constants are calculated as $k_2 = E_2/H$ and $k_4 = E_4/H^3$, where the elastic moduli $E_2$ = 1.1 MPa and $E_4$ = 2.9

MPa are fitted from the experimental data (Ref. [4]), as shown in FIG. S1. With these materials parameters and taking $\sigma_0 = 1$ MPa, for instance, the characteristic wavenumber $q = 0.21$ μm$^{-1}$, the dimensionless quantities $\lambda = 1.1 \times 10^{-6}$ and $\varepsilon = 0.84$. Under ambient pressure ($p \approx 0.1$ MPa), the mean value of the dimensionless deflection $\psi$ is estimated to be $\bar{\psi} \approx -qH(p/E_2) \approx -58$.

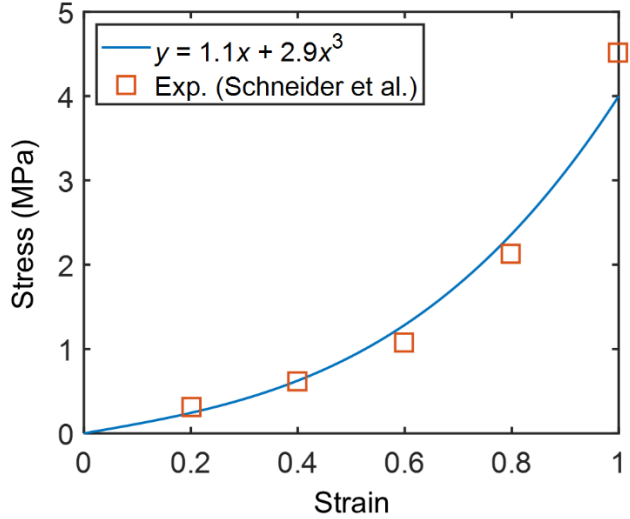

FIG. S1. Spring constants calibrated from the experimental stress-strain curve of PDMS (Sylgard 184, Dow Corning).

For simulations under constant compressive stress, the $\psi$ field is drawn from $t = 2000$ where the pattern has been stabilized. $128 \times 128$ square mesh grids are used. For the case of cyclic loading, a square waveform is adopted with a period of 4000. $256 \times 256$ mesh grids are used. An implicit method is adopted for time stepping with MUMPS solver in all these simulations.